\documentclass[twocolumn,floatfix,prl]{revtex4}
\begin{document}

\title{Comment on \\
``Dynamic Wetting by liquids of different viscosity'', by \\
   T.D. Blake and Y.D. Shikhmurzaev
            }
\author{Jens Eggers$^{\dagger}$  and Robert Evans$^{\ddagger}$ 
            }
\affiliation{
$^{\dagger}$ School of Mathematics, 
University of Bristol, University Walk, \\
Bristol BS8 1TW, United Kingdom  \\
$^{\ddagger}$ H.H. Wills Physics Laboratory, University of Bristol \\
Bristol BS8 1TL, United Kingdom
        }
\begin{abstract}
We comment on a recent theory of dynamic wetting, that is based 
directly upon a model for interface formation, introduced by
Shikhmurzaev. We argue that the treatment of surface tension
and its relaxation, inherent in the original model, is physically flawed. 
\end{abstract}

\maketitle
In a recent paper \cite{BS02}, Blake and Shikhmurzaev utilize
a model for interface formation, proposed initially by 
Shikhmurzaev \cite{S93,S94}, to interpret their measurements of 
the velocity dependence of the dynamic contact angle at a moving 
contact line. They determine some of the parameters 
contained in the model, in particular, their viscosity 
dependence. In this comment, we point out that the key
quantities upon which this model is based have no well-defined 
physical meaning. More specifically, the treatment of surface 
tension which underlies the theory does not correspond to any 
known physical mechanism. As a result, the model leads to 
consequences which are physically absurd and which are at odds 
with available experimental data. Furthermore, we show that 
the values for model parameters, purportedly estimated in \cite{BS02},
depart by many orders of magnitude from what is physically
reasonable. 

The model in Ref.\cite{BS02} is based on the fundamental 
assumption that the surface tension of a {\it pure liquid} is 
determined by a `surface equation of state', which 
fixes its value as function of surface parameters (see also \cite{S93}, 
section 2.1). As an approximation, the surface tension $\sigma$ is assumed 
to depend on the `surface density' $\rho^s$ alone, cf. equations (1)
and (4) of \cite{BS02}:
\begin{equation}
\label{sigma}
\sigma = \gamma(\rho_0^s - \rho^s),
\end{equation}
where $\gamma$ and $\rho_0^s$ are phenomenological constants. 
In \cite{BS02}, Shikhmurzaev's model is applied to both the free
liquid-gas and the liquid-solid interface, but we will focus on 
the former, so there is no need for an index on $\sigma$. 
Equation (\ref{sigma}) makes no physical sense. Firstly, as 
explained in \cite{RW82}, p. 31 ff, surface density as introduced 
by Gibbs cannot be an intrinsic property of the surface
of a pure liquid. It depends solely on the definition of what 
Gibbs calls the {\it dividing surface} between the two phases. 
Conventionally, for a pure liquid, one defines the dividing surface in such
a way that the surface density {\it vanishes} (\cite{RW82}, p. 31),
a choice which is called the {\it equimolar surface}. 
Secondly, the surface is not an independent thermodynamic system
that would allow relations between its extensive and intensive 
parameters to be defined, as is done in \cite{S93,S94,BS02}, and as 
is implied by (\ref{sigma}). As  pointed out in \cite{RW82}, p.33, 
the surface exists only by virtue of the bulk phases that surround it;
it does not form an autonomous phase. Hence an expression such as
$\sigma=\sigma(\rho^s)$, as given in \cite{BS02}, is meaningless
as regards surface thermodynamics, no matter how $\rho^s$ is defined.

We illustrate our criticism further by pointing out two of the 
consequences of the ansatz (\ref{sigma}), which we think are absurd. 
Equation (13) of \cite{BS02} states that the surface tension should go to 
zero with surface thickness $h$ (assuming $\rho_e^{s*}$ is constant, 
as is implied in the text). This is well known not to be the case,
see \cite{RW82} p. 16 ff and p.47, which provides an authoritative 
critique of treatments based on point-thermodynamics. Of course, 
$h$ is not really an independent quantity, rather, it is determined 
by the conditions of thermodynamic equilibrium at the surface. 
However by considering, for example, different temperatures, 
different interface density profiles can effectively
be realized. Given the interface profile, the surface tension 
can be calculated essentially by mechanical arguments \cite{RW82}.
(For a slowly-varying profile, treated in the square-gradient 
approximation of Rayleigh and van der Waals, the result is equation 
(1.43) or (3.11) of \cite{RW82}.) If anything, the surface tension 
will be {\it larger} for a sharp interface ($h=0$) then for the 
real smooth one. The reason is that the real interface shape 
is one that {\it minimizes} the total free energy (grand potential)
of the inhomogeneous fluid (see \cite{RW82}, p. 54 ff). Indeed, at 
high temperatures, when the interface becomes more diffuse, 
surface tension {\it decreases} (cf. Figs. 1.5, 1.6, and 6.5 \cite{RW82}). 

Next, we consider equation (14) of \cite{BS02} for the typical time
scale $\tau$, over which the surface tension reaches its equilibrium 
value after a fresh surface is created. The time $\tau$ is claimed 
to be proportional to the viscosity $\mu$ of the fluid, and a specific 
estimate is given for $\mu=672 mPa s$, for which the value is supposed 
to lie between $\tau=2.5\times10^{-6}s$ and $8.3\times10^{-6}s$.
To a first approximation, we in fact believe that the timescale 
needed to establish a surface tension in a pure fluid is essentially 
zero: force is transmitted with the speed of light, giving 
$\tau\approx 10^{-18}s$ for molecular sizes. The interfacial profile, 
and thus the surface tension, subsequently relaxes toward its equilibrium 
value, but even for this process the author's numbers appear to be a 
gross over-estimation. The thickness of the liquid-gas interface 
is typically 2 or 3 molecular diameters near the triple-point 
(\cite{RW82}, Chs 6 and 7); this gives about $5\times 10^{-10} m$. Dividing 
this length by the speed of sound in a typical simple liquid, we arrive at 
$\tau_s\approx  10^{-12} s$. In other words, $\tau$ will be
given by a typical collision time in the liquid, which is not
related in any simple way to the fluid viscosity. 

This estimate is 
consistent with typical decay times of correlation functions in
simple liquids measured by neutron scattering as well as with those found in
molecular dynamics simulations \cite{HM86}. Note also that Brillouin 
scattering from density fluctuations in bulk liquids, where the 
wavelengths involved are much larger,of the order of $5\times10^{-7} m$,
is characterized by shifts of frequency of
typically $10^{10} Hz$ and that Brillouin spectra for liquid interfaces
correspond to the same range of frequency shifts \cite{D82}. 
We are not aware of any
physical mechanism that would give rise to relaxation times in the order 
of $10^{-6}s$.

One might argue that as the authors' estimates are based on their own
measurements of a moving contact line, they should provide independent 
evidence for the consistency of the assumptions of the model. Unfortunately, 
we do not believe this to be the case. The reason is that the authors'
theory refers to what they term the dynamic contact angle $\theta_d$, 
determined solely by a balance of surface tensions, cf. equation (7)
of \cite{BS02}.
Viscous forces and, therefore, interface bending does not enter their
description. The {\it measurement} of the interface angle was
however performed at a scale of about a $mm$. It is well known that 
the interface near a moving contact line is highly curved \cite{G85},
which is the result of viscous forces which therefore cannot be ignored. 
This is best appreciated in 
the case of a perfectly wetting fluid, where the contact line is 
preceded by a precursor film \cite{G85}. Hence no interface formation 
is taking place, yet on macroscopic scales measured contact angles 
have a speed dependence consistent with Tanner's law \cite{G85}. 
This implies that in the partially wetting case, considered here, 
any effects of interface bending would have to be carefully 
subtracted for a correct interpretation of experimental data.


\begin{thebibliography}{0}
\bibitem{BS02}
T.D. Blake and Y.D. Shikhmurzaev, 
J. Coll. Interf. Sci. {\bf 253}, 196 (2002). 

\bibitem{S93}
Y.D. Shikhmurzaev, 
Int. J. Multiphase Flow {\bf 19}, 589 (1993). 

\bibitem{S94}
Y.D. Shikhmurzaev, 
Fluid Dynamics Research {\bf 13}, 45 (1994). 

\bibitem{RW82}
J. S. Rowlinson and B. Widom, {\it Molecular theory of capillarity},
Oxford, 1982.

\bibitem{HM86}
J.P. Hansen and I.R. McDonald, {\it Theory of Simple Liquids, 2nd Edition},
Academic Press, 1986.

\bibitem{D82} 
J.G. Dil,
Rep. Progr. Phys. {\bf 45}, 285 (1982).

\bibitem{G85} 
P. G. de Gennes,
Rev. Mod. Phys. {\bf 57}, 827 (1985).

\end{thebibliography}
\end{document}